\documentstyle[aps, pra, tighten]{revtex}
\begin{document}
\draft
\hoffset= -2.5mm

\title{On the `Anomalous' Resurgence of Shot Noise in Long Conductors}

\author{Mukunda P. Das \\}
\address{
Department of Theoretical Physics,
Research School of Physical Sciences and Engineering, \\
The Australian National University,
Canberra ACT 0200, Australia \\}

\author{Frederick Green \\}
\address{
GaAs IC Prototyping Facility Program,
CSIRO Telecommunications and Industrial Physics,\\
PO Box 76, Epping NSW 1710, Australia \\}

\maketitle

\begin{abstract}
There has been renewed interest in the physics of the so-called
`crossover' for current fluctuations in mesoscopic conductors,
most recently involving the possibility of its appearance
in the passage to the macroscopic limit.
Shot noise is normally absent from solid-state conductors
in the large, and its anomalous resurgence
there has been ascribed to a rich interplay
of drift, diffusion, and Coulomb screening.
We demonstrate that essentially the same rise
in shot noise occurs in a much less complex system:
the Boltzmann-Drude-Lorentz model of a macroscopic,
uniform gas of {\it strictly non-interacting} carriers.
We conclude that the `anomalous crossover' is a
manifestation of simple kinetics.
Poissonian carriers, if driven by a {\it high enough field},
cross the sample faster than any scattering time,
thus fulfilling Schottky's condition for ideal shot noise.

\end{abstract}


\section{Introduction}

The fine-scale investigation of carrier noise in mesoscopic
conductors is now a well-established field within the transport
physics of the solid state. It has reached new heights
recently, experimentally and theoretically
%\cite{blbu}.
(Blanter and B\"uttiker 2000).
Noise is a unique source of information on the dynamics of
microscopic fluctuations. This is notably so at small
length scales, already approaching the quantum domain
in actual devices.

A particular aspect of mesoscopic noise, in metallic diffusive
conductors especially, is the so-called `crossover'
from thermal to shot noise.
There are many measurements of it, and almost as many compelling
(if frequently quite disparate) theoretical explanations.
The term {\it crossover} refers to the smooth evolution, with
increasing voltage, of the current-noise spectral density $S(V;\omega)$
(normally it is sufficient to study
its low-frequency limit $\omega \ll \tau^{-1}$, where
$\tau$ is a characteristic collision time).
One sees the onset of a non-dissipative excess
component in $S(V;0)$, over and above the dissipative
Johnson-Nyquist noise which exhausts the low-field limit.
Thus, typically,

\begin{equation}
{S(V;0)\over S_0}
= 1 + \gamma {\left[ {eV\over 2k_{\rm B}T}
\coth{\left( {eV\over 2k_{\rm B}T} \right)}  - 1 \right]}
\label{eq1}
\end{equation}

\noindent
where $S_0 = 4Gk_{\rm B}T$ is the Johnson-Nyquist value
and $\gamma$ the suppression factor; $G$ is the sample conductance
and $k_{\rm B}T$ the thermal energy.
The factor $\gamma$ is a signature of the often
subtle correlation effects in the microscopic fluctuations,
which are responsible for the form of $S(V;0)$ as measured
%\cite{blbu}.
(Blanter and B\"uttiker 2000).
At voltages $V \gg k_{\rm B}T/e$, Equation (\ref{eq1})
gives $S(V;0) = 2\gamma eI$, where $I = GV$ is the current.
This exhibits suppression of the Schottky formula
$S = 2eI$ for classical Poissonian shot noise.

While Eq. (\ref{eq1}) gives an impressive empirical fit to many
(though not all) experiments, serious questions arise
as to whether any of the prevailing models
%\cite{blbu,kogancpu}
(Kogan 1996; Blanter and B\"uttiker 2000)
possesses the internal consistency expected
of a standard microscopic description. 
%We cite References \onlinecite{ithaca} and \onlinecite{upon}
We cite Das and Green (2000) and Green and Das (2000{\it a})
for a description of what may go awry with such theories,
all of which rely heavily on hydrodynamic drift-diffusion
analogies for mesoscopic transport
%\cite{kogancpu}.
(Datta 1995; Kogan 1996).
Theoretically, the concept of the `crossover' may be as open
to new critiques (Gillespie 2000)
as it is to new empirical tests by appropriately
designed experiments
%\cite{upon,gdcond,gd01,gdii}.
(Green and Das 1998{\it a}, 2000{\it a-c}).

Now, a fresh window on the `crossover' appears to have
%been opened by Gomila and Reggiani's recent work
%\cite{gomila}.
been opened by the recent work of Gomila and Reggiani (2000).
Foremost is their emphasis on the explicit role of
carrier-number fluctuations as generators of the observable
shot noise. We note that this perspective was addressed
theoretically (with detailed computations) in
%Ref.
%\onlinecite{gdcond}.
%See also Ref.
%\onlinecite{upon}.
Green and Das (1998{\it a}, 2000{\it a}).

There is a basic difference between fluctuations of carrier number,
manifesting at the conductor-lead interfaces and engendering shot noise,
and fluctuations of the free energy, manifesting throughout
the conductor's volume and generating thermal noise.
This understanding is in sharp contrast to the
usual phenomenological viewpoint
%\cite{kogancpu}
(Kogan 1996)
in which no difference is permitted, {\it even in principle},
between thermally related and carrier-number related processes.
In this respect it is useful to bring to mind the textbook
distinction between a variation with respect to chemical
potential and a variation with respect to
particle number. The fact that they are intimately
linked by microscopics does not override the fact
that they are thermodynamic conjugates,
with wholly distinguishable thermodynamic consequences.

Gomila and Reggiani (2000)
certainly raise weighty, if not wholly unanticipated,
points regarding macroscopic solid-state shot noise
(Green and Das 1998{\it a}; Naveh 1998).
Furthermore, these are not easily addressed in
strictly low-field, linear, drift-diffusive descriptions
%\cite{blbu,kogancpu,datta}.
(Datta 1995; Kogan 1996; Blanter and B\"uttiker 2000).
Thus it is well to revisit our own existing non-perturbative
Boltzmann theory, allied to a time-of-flight intepretation
of shot noise
%\cite{gdcond}.
(Green and Das 1998{\it a}).
In Section II we briefly recall
our formalism, while in Sec. III we give simple Drude-like
(but complete and exact) kinetic solutions for
shot noise as a function of current, dimensionality,
length, and finite radius of the conductor.
These show that the `crossover' anomaly exists
with no reference {\it at all} to long-range Coulomb correlations.
In Sec. IV we examine the results of
%Ref.
%\onlinecite{gomila},
Gomila and Reggiani (2000),
based on a diffusive Langevin-Poisson scheme,
and compare them with those of
%Ref.
%\onlinecite{gdcond}
Green and Das (1998{\it a})
as recalled in the present paper.
Section V contains our final remarks.

\section{Kinetics of shot noise: theory}

\subsection{Non-equilibrium fluctuations}

In the context of
%Ref.
%\onlinecite{gomila}
Gomila and Reggiani (2000)
we specialise to a homogeneous
metallic wire, subject to a uniform driving field.
The mean carrier density is $n$ and the mean
total carrier number is $N = \Omega n$ in the sample volume $\Omega$.
These are independent of the external field. The kinetic Boltzmann
equation for the spatially uniform distribution
function $f_{\bf k}(t)$ is, in the Drude-Lorentz collision approximation,

\begin{equation}
{\left( { {\partial }\over {\partial t} } 
+  {
   { { eE}\over {\hbar} }
  }
   { {\partial }\over {\partial k_x} }
\right)} f_{\bf k}(t)
= - { 1\over \tau}
{\left( f_{\bf k}(t)
- { {\langle f(t) \rangle} \over {\langle f^{\rm eq} \rangle} }
f^{\rm eq}_{\bf k} \right)}.
\label{eq2}
\end{equation}

\noindent
Our electrons are positive for convenience.
The field $E$ is in the source-drain $x$-direction, and $\tau$
is the collision time. We denote traces over wave-vector
(with a factor of two for spin) as
${\langle f \rangle} \equiv 2\Omega^{-1}\sum_{\bf k} f_{\bf k}$.
The physical constraint on the traces in the right side of Eq. (\ref{eq2})
is, naturally,
${\langle f(t) \rangle} = {\langle f^{\rm eq} \rangle} = n$.
Finally,

\[
f^{\rm eq}_{\bf k}
= [1 + \exp(\varepsilon_{\bf k} - \mu)/k_{\rm B}T)]^{-1}
\]

\noindent
is the usual Fermi-Dirac equilibrium distribution, parametrised
by the thermal energy and by $\mu$, the chemical potential.
Note that there is no coupling to the Poisson equation,
since the system is uniform.

To determine all the relevant microscopic correlation
functions in this non-equilibrium system, one must generate the
distribution of its {\it electron-hole pair} fluctuations.
That requires variational analysis of Eq. (\ref{eq2}).
There are two steps in this, related but treated separately.
First we compute the steady-state fluctuation distribution

\begin{equation}
\Delta f_{\bf k} \equiv k_{\rm B}T
{{\delta f_{\bf k}}\over {\delta \mu}}
= {1\over \Omega} \sum_{\bf k'}
{{\delta f_{\bf k}}\over {\delta f^{\rm eq}_{\bf k'}}}
\Delta f^{\rm eq}_{\bf k'},
\label{eq3}
\end{equation}

\noindent
where the equilibrium fluctuation is the mean-square
fluctuation of the occupation number,
{\it precisely} as defined in statistical mechanics:

\begin{equation}
\Delta f^{\rm eq}_{\bf k} = k_{\rm B}T
{{\delta f^{\rm eq}_{\bf k}}\over {\delta \mu}}
= f^{\rm eq}_{\bf k} (1 - f^{\rm eq}_{\bf k}).
\label{eq4}
\end{equation}

\noindent
Equation (\ref{eq3}) is directly calculable by variation
of the one-body Boltzmann transport equation (BTE),
and is the exact solution to the linearised BTE
(Green and Das 2000{\it b,c}).

For every collision model, there exists a unique
one-to-one transformation that maps $\Delta f^{\rm eq}$
to the functional $\Delta f$ in the steady state.
Together with the fact that $\Delta f$ exactly satisfies
the variational BTE for the electron-hole pair
(density-density) correlation function
(Kadanoff and Baym 1962), such a mapping establishes
Eq. (\ref{eq3}) as the unique mathematical form
of the mean-square particle fluctuations out of equilibrium.
This explicit and crucial connection, between the equilibrium
and non-equilibrium fluctuations, is not model-dependent.
It is completely generic to the kinetic description of noise
and completely absent from every drift-diffusive description
(Kogan 1996).

The next step is to obtain the dynamic response. In our physical
picture, a spontaneous thermal energy exchange with the
heat reservoir sets up an initial electron-hole pair excitation with
average strength $\Delta f_{\bf k'}$, at some given initial
location ${\bf r'}$ within volume $\Omega$. The localised
spontaneous excitation is not stable. It relaxes back to the
steady state according to the
time-dependent Green function, derived when Eq. (\ref{eq2})
is perturbed in time, wave-vector space, and (weakly) in real space.
This dictates the change in background distribution
for state ${\bf k}$ at position ${\bf r}$,
given the initial random excitation at ${\bf r'}$.
Full technical details are in
%Refs.
%\onlinecite{gdcond,gd01,gdii}.
Green and Das (1998{\it a}, 2000{\it b,c}).

All of the relaxation dynamics
are contained in the {\it transient} component of the calculable
Green function, remaining after the stable long-time adiabatic part
is removed from the full response.
Let us denote the transient component, in the frequency domain,
by ${\cal C}_{{\bf k} {\bf k'}}({\bf r} - {\bf r'}; \omega)$.
The dynamic (and non-local) two-body response,
namely the electron coupled with its hole, is denoted by

\begin{equation}
\Delta {\rm f}^{(2)}_{{\bf k} {\bf k'}}({\bf r} - {\bf r'}; \omega)
\equiv {\cal C}_{{\bf k} {\bf k'}}({\bf r} - {\bf r'}; \omega)
\Delta f_{\bf k'}.
\label{eq5}
\end{equation}

\noindent
The full velocity-velocity correlation function is,
following the approach of Gantsevich {\it et al.}
%\cite{ggk},
(1979)

\begin{equation}
{ \langle\!\langle {\bf v} {\bf v'}
{\Delta {\rm f}^{(2)} } ({\bf r} - {\bf r'}; \omega) 
\rangle\!\rangle}_c'
\buildrel \rm def \over =
{2\over \Omega^2}
\sum_{\bf k} \sum_{\bf k'}
{\bf v}_{\bf k}
{{\rm Re} \{{\cal C}_{{\bf k} {\bf k'}}({\bf r} - {\bf r'}; \omega)\} } 
{\bf v}_{\bf k'} \Delta f_{\bf k'}.
\label{eq6}
\end{equation}

\noindent
This completely determines the response of the carrier flux
at ${\bf r}$, induced by a spontaneous thermal fluctuation
in the carrier flux at ${\bf r'}$. This
is {\it not} a reciprocal process; although uniformity means that the
magnitude of the correlator depends on the relative
co-ordinate ${\bf r} - {\bf r'}$, it matters which of the
two positions is upstream. The externally driven system fluctuates
asymmetrically in space, just as it fluctuates irreversibly in time.

We now have the basic tool to construct both the thermal noise
and the shot noise in the conductor.
We discuss the low-frequency case.
The thermal spectral density becomes, in a familiar way,
the volume integral of all the current-current correlations:

\begin{equation}
S_{\rm therm}(E)
\equiv 4 \int_{\Omega} d{\bf r} \int_{\Omega} d{\bf r'}
{ \langle\!\langle (ev_x/L) (ev'_x/L)
{\Delta {\rm f}^{(2)} } ({\bf r} - {\bf r'}; 0) 
\rangle\!\rangle}_c';
\label{eq7}
\end{equation}

\noindent
the conductor's length is $L$.
The expression is often portrayed in the literature
as a real-space symmetrised form which, however,
adds little or nothing to the intrinsic physics.
We do not elaborate on $S_{\rm therm}(E)$ except to recall two
major properties. The first is the Johnson-Nyquist equilibrium limit

\begin{equation}
S_{\rm therm}(E \to 0) = 4Gk_{\rm B}T
\label{eq8}
\end{equation}

\noindent
where, in our Drude model, the conductance becomes
$G = N e^2 \tau/m^*L^2$, for effective mass $m^*$.

The second property is that Eq. (\ref{eq7}), regardless
of driving field, {\it will never scale other than as}
$\Delta f \sim \Delta f^{\rm eq} \sim T$
{\it in a degenerate metallic conductor}.
This is the strict and inevitable
consequence of kinetics, of Fermi-liquid physics,
and most of all of asymptotic equilibrium
and neutrality in the metallic leads.
It has been discussed exhaustively
%\cite{ithaca,upon,gdcond,gd01,gdii}.
(Das and Green 2000; Green and Das 1998{\it a}, 2000{\it a-c}).

\subsection{Shot noise}

To set the work of Gomila and Reggiani in context, we must
look first at the `smooth crossover formula' and its
inbuilt theoretical deficiency.
True shot noise never scales with temperature; for example,
it remains well-defined even in the zero-temperature limit.
In purporting to make thermal noise integral with true shot noise,
the `smooth-crossover formula' of
drift-diffusive theory, Eq. (\ref{eq1}),
attempts the kinetically impossible.
For, every drift-diffusive phenomenology proclaims that
Eqs. (\ref{eq1}) and (\ref{eq7}) are identical
(Kogan 1996; Blanter and B\"uttiker 2000).
Yet the rigorous kinetic-theoretical constraint on
non-equilibrium thermal noise,
namely its abiding proportionality to $T$,
means that Eq. (\ref{eq7}) {\it cannot possibly describe shot noise}
in the presence of strong degeneracy.

It follows that Eq. (\ref{eq1}) is hard to justify.
At any rate, the equation's theoretical claim
(thermal noise equals shot noise)
is unsustainable by a first-principles analysis.
We mean an analysis that is conventionally executed,
in keeping with the conventional understanding of
statistical mechanics and microscopics
(Green and Das 2000{\it a,b}).
The `smooth crossover formula' is
inconsistent, purely and simply.

That is the background to the shot-noise
considerations of Gomila and Reggiani (2000).
We do not deny that shot-noise-{\it like}
structure -- linear in the current -- can emerge
from the thermal noise spectrum.
Indeed it does, in the semiclassical ballistic limit
(Green and Das 1998{\it b}; Gomila and Reggiani 2000).
It is also the case that purely classical models
lead to classical shot noise, $2eI$,
based on Eq. (\ref{eq7}).
Nevertheless the leading concern is with
{\it metallic} diffusive wires. There, $T$-independent
shot noise cannot be contrived from a strictly thermal basis.
Detailed experiments have been proposed to test our claim
(Green and Das 2000{\it a}, 2000{\it c}).

With the knowledge that true shot noise is fundamentally different
from thermal noise, one can build an operationally
consistent theory for it. We introduce the idea of
the response to variations
in the total number of carriers transiting the
conductor. In the mean, $N$ is constant in time but
fluctuates by $\delta N^+ = +1$ at any instant that
a carrier {\it first enters at the source}. Similarly it
changes by $\delta N^- = -1$ as the carrier
{\it finally exits at the drain} (conceptually this
amounts to the injection of a hole at the drain).
It is not hard to see that this process is described
by the non-local correlation

\begin{equation}
{\cal C}_{{\bf k} {\bf k'}}({\bf r} - {\bf r'}; 0)
{{\delta f_{\bf k'}}\over {\delta N}} \delta N^{\pm}
= {\cal C}_{{\bf k} {\bf k'}}({\bf r} - {\bf r'}; 0)
{{\Delta f_{\bf k'}}\over {\Delta N}} \delta N^{\pm},
\label{eq9}
\end{equation}

\noindent
where $\Delta N = \Omega {\langle \Delta f \rangle}$.
When this correlation records a particle entry, then ${\bf r'}$
lies in the cross-sectional region at the source and ${\bf r}$ at
the drain. When it records a particle exit, then ${\bf r'}$
belongs to the drain area and ${\bf r}$ to the source.
Figure 1 illustrates the principle. The essence of shot noise is
that it is a time-of-flight process, involving many sporadic transits
of carriers across a predefined geometry. On this view
shot noise has nothing to do with correlations distributed
throughout the volume of a conductor. This is totally unlike
the guiding assumption for Eq. (\ref{eq1})
%\cite{kogancpu}.
(Kogan 1996).

The definition of the measurable shot noise, strictly across the
source-drain gap, follows naturally. It simply sums the stochastically
independent terms produced, on average,
by each of the $N$ active contributors.
Each contribution is of equal weight since the carriers are all
equivalent (assuming temporal randomness of entry/exit).
Thus

\begin{mathletters}
\label{eq10}
\begin{equation}
S_{\rm shot}(E)
\buildrel \rm def \over = 
S_{\rm s;d}(E)\delta N^+ + S_{\rm d;s}(E)\delta N^-
= S_{\rm s;d}(E) - S_{\rm d;s}(E),
\label{eq10a}
\end{equation}

\noindent
in which the two directional correlations are

\begin{equation}
S_{\rm s;d}(E)
= 2N \int d{\bf r} \delta(x - L) \int d{\bf r'} \delta(x')
{1\over \Delta N}
{ \langle\!\langle (ev_x) (ev'_x)
{\Delta {\rm f}^{(2)} } ({\bf r} - {\bf r'}; 0) 
\rangle\!\rangle}_c'
\label{eq10b}
\end{equation}

\noindent
for injection at the source $(x = 0)$ and, for
removal at the drain $(x = L)$, 

\begin{equation}
S_{\rm d;s}(E)
= 2N \int d{\bf r} \delta(x - L) \int d{\bf r'} \delta(x')
{1\over \Delta N}
{ \langle\!\langle (ev_x) (ev'_x)
{\Delta {\rm f}^{(2)} } ({\bf r'} - {\bf r}; 0) 
\rangle\!\rangle}_c'.
\label{eq10c}
\end{equation}

\end{mathletters}

\noindent
Note especially that:
\begin{itemize}
\item
the space co-ordinates in the argument
of $\Delta {\rm f}^{(2)}_{{\bf k} {\bf k'}}({\bf r'} - {\bf r}; 0)$
are reversed in $S_{\rm d;s}(E)$, and
\item
$S_{\rm shot}(E = 0)$ vanishes, since the equilibrium
kinetic equation is self-adjoint (time reversible)
and entails the identity $S_{\rm s;d}(0) = S_{\rm d;s}(0)$.
\end{itemize}

One easily verifies that Eq. (\ref{eq10}) is
explicitly independent of temperature
and goes to $2eI$ for current $I$ in the
semi-classical ballistic limit
%\cite{gdcond}.
(Green and Das 1998{\it a}).
Neverthless, the intimate microscopic link with the
(conjugate) thermal effects remains. It is manifest in the role
of the flux auto-correlation
${ \langle\!\langle v_x v'_x
{\Delta {\rm f}^{(2)} } ({\bf r} - {\bf r'}; 0) 
\rangle\!\rangle}_c'$.

Ultimately, a real measurement of current fluctuations
detects them in the access leads for the sample. Hence
we expect to detect the sum of thermal and shot-noise
contributions, if these effects are statistically independent.
Non-equilibrium thermal noise will itself carry a hot-electron
excess; it is non-dissipative rather than Johnson-Nyquist in origin
%\cite{gd01,gdii},
(Green and Das 2000{\it b,c}),
just as shot noise is non-dissipative.
In the macroscopic limit
hot-electron noise goes quadratically with $E$, at least
in simple cases.
One should ask whether this term could overwhelm an
emerging shot-noise signal. This is unlikely,
as can be seen from a rough estimate
based on the Drude model. The bulk thermal-noise
excess goes as

\begin{equation}
S_{\rm exs}(E) = S_0 {{\Delta N}\over N}
{{m^* \mu_{\rm e}^2 E^2} \over k_{\rm B}T} \leq 
4G m^* \mu_{\rm e}^2 E^2
\label{eq11}
\end{equation}

\noindent
since the ratio $\Delta N/N$ is always less than one
in a degenerate conductor.
(Here $\mu_{\rm e} = e\tau/m^*$ is the mobility.)
By finding the upper bound to $2eI = 2e(GEL) \geq S_{\rm exs}(E)$
for typical material parameters, one concludes that pure
shot noise -- if there were no other mechanism to suppress it -- would
dominate at least up to fields $\sim 10^6{~}{\rm Vcm}^{-1}$
for a sample 1 mm long. At much shorter (mesoscopic) lengths
this simple estimate does not hold; specific modelling is needed.

\section{Kinetics of shot noise: application}

We can now review our results from
%Ref.
%\onlinecite{gdcond},
Green and Das (1998{\it a}),
built on the form of $S_{\rm shot}(E)$.
Despite the relatively crude form of the inelastic
collision term in the Drude model, one might expect it to
be more relevant at {\it high fields} than,
say, linear treatments that emphasise coherent (or at least elastic)
scattering. A driving potential of a just few
volts is quite enough to place a conductor,
some millimetres in length, beyond the validity of
purely elastic scattering adrift-diffusive theory
(Green and Das 2000{\it b}).

The calculation is straightforward. The functions
$f_{\bf k}$ and $\Delta f_{\bf k}$ are first obtained
for a given field $E = V/L$. Then the two-point transient response
${\cal C}_{{\bf k} {\bf k'}}({\bf r} - {\bf r'}; \omega=0)$
is derived from the linearised BTE.
Finally, all are combined to yield Eq. (\ref{eq10}).

In Fig. 2 we plot the sum of thermal and shot noises in
a one-dimensional (1D) wire calculated within the Boltzmann-Drude model
of transport, Eq. (\ref{eq2}). This is in the strongly
degenerate carrier regime,
at a thermal energy chosen as $k_{\rm B}T = 0.1\varepsilon_{\rm F}$,
with Fermi energy $\varepsilon_{\rm F}$. The spectral density is normalised
to the Johnson-Nyquist value $S_0$ and displayed as a function
of current in units of $I_0 = 2Gk_{\rm B}T/e$.
In this and subsequent figures, each curve corresponds to one of
five values of the device length as a ratio with the mean
free path $\lambda = \tau v_{\rm F}$ in terms of the Fermi velocity.
The curves are always monotonic, tracking down as the ratios
rise in the sequence
$L/\lambda = 0.001, 1, 10, 25, {~}{\rm and}{~} 50$.
Typically, $\lambda$ is of the order of 50 to 100 nm.

The shortest wire exhibits full shot noise. This is
the semi-classical ballistic limit: the absolute
upper bound for our model.
In the case of the second shortest wire $L = \lambda$,
the curve falls a little below
the ideal value $1 + 2eI/S_0$ in the topmost curve.
We note a slight shoulder at $I \approx 2G\varepsilon_{\rm F}/e$.
The shot noise remains quasi-ballistic because Pauli blocking in
the 1D free electron gas efficiently inhibits any scattering
when carriers cannot gain enough energy to leave the Fermi sea.
This makes the Fermi distribution fairly robust
to moderate external fields.

In longer wires, the shot noise at moderate currents is
attenuated more and more. Inelastic-scattering
suppression is exponential in the Drude model,
taking the low-current form

\begin{equation}
{S_{\rm shot}(E)\over 2eI}
\to {\left( 1 +
{L\over \lambda} + {L^2\over 2\lambda^2} 
\right)} e^{-L/\lambda},
\label{eq12}
\end{equation}

\noindent
which dies very quickly as the length increases.
This accounts for the macroscopic extinction of shot noise.

Figure 3 displays the shot noise of classical 1D carriers, with
the same $n$ and $T$ as the degenerate system of Fig. 2.
The mean free path is now $\tau v_{\rm th}$ where
$v_{\rm th} = (2k_{\rm B}T/m^*)^{1\over 2}$. We have retained
the same physical wire lengths $L$ as in Fig. 2
so that, for Fig. 3 specifically, our chosen ratios of
length to mean free path are scaled up by 
$v_{\rm F}/v_{\rm th}$.
It is significant that the classical curves fall mostly
on top of the degenerate ones. An obvious exception is the second
plot, where we saw that degeneracy shields the shot noise
from attenuation at lower $I$.
Otherwise the high-current behaviour of the 1D shot noise
is unchanged, and thus independent of the carrier statistics.
From this it is evident that degeneracy plays no role
in the resurgence of high-field shot noise.

To round off our discussion we compare the 1D case with the
three-dimensional (3D) case. This is of interest (i) because
it is closer to actual experimental devices, and (ii) because
it allows us to study the effect of finite, even narrow, wires.
Similar results apply in two dimensions (Green and Das 1998{\it a}).

Fig. 4 shows spectra for the same sequence of lengths,
with a wire radius $R = 100 \lambda$; very wide.
There are minor differences with Fig. 2. For instance,
in the second curve Pauli blocking is somewhat less
effective in preventing attenuation of low-current shot noise.

Rather more interesting are Figs. 5a and 5b. The first is for
a wire radius $R = 0.3 \lambda$
and the second for $R = 0.05 \lambda$.
(For $\lambda \sim$ 100nm, such thicknesses should be
achievable by sophisticated nano-lithography.)
There is a major loss of shot-noise spectral strength
relative to Fig. 4.
Recovery towards full noise does not occur before considerably
higher current levels are reached.
The freedom to explore shot noise through
a new variable, the thickness, suggests a novel
range of experiments to complement those proposed
for shot-noise resurgence as a function of length
%\cite{gomila}.
(Gomila and Reggiani 2000).
Such experiments will in any case require exploration
of intrinsically {\it high-field}, {\it high-current} behaviour.
This is a region dominated by inelastic collisions,
and one that has been almost totally
neglected in metallic mesoscopic systems.

Space prevents an extended presentation of our work.
We invite readers to examine the complete exposition
of our general kinetic approach to noise in
%Refs.
%\onlinecite{gd01}
%and
%\onlinecite{gdii}
Green and Das (2000{\it b,c})
as well as the specifics of shot noise in
%Ref.
%\onlinecite{gdcond}.
Green and Das (1998{\it a}).

Before discussing the work of Gomila and Reggiani
(2000),
we remark on the striking graphical similarity between their
calculation of $S_{\rm shot}(E)$ and the present one.
On the other hand, Gomila and Reggiani's theory
for the anomalous rise of macroscopic shot noise apparently
depends, in large degree, on the role of long-range Coulomb
correlations. In our results, Coulomb effects
are completely absent in the sample.
Nevertheless, the behaviour
of the shot noise is practically the same in our
inelastic free-electron model.

\section{The Gomila-Reggiani theory}

The recent paper of Gomila and Reggiani
%\cite{gomila}
(2000)
presents a much-needed invitation to reassess
the microscopic basis of shot noise
(unavoidably, this brings in the questionable status
of the `smooth crossover').
Certainly Gomila and Reggiani
make their argument on the overriding idea of shot noise
as a number-fluctuation phenomenon, much in the
way of our own philosophy, as already discussed
elsewhere
%\cite{gdcond,upon}.
(Green and Das 1998{\it a}, 2000{\it a}).
However, there are some differences between the respective
approaches. For example, Gomila and Reggiani propose a
drift-diffusive model. This entails additional, intuitive
assumptions meant to simplify the underlying transport problem.
By contrast, we work directly with the semi-classical kinetic equation.
The main points of difference are:

\begin{itemize}
\item{\bf Diffusive method}.
The drift-diffusive equation of motion embodies
a model for current fluctuations in which the diffusion constant $D$
(in microscopic terms, a current-current correlator)
appears as a simple scaling parameter for the evolution of
the current fluctuations themselves. This means that the solution
preconditions its own scaling factor $D$. If one stays
rigidly within the linear-response limit,
the Einstein relation can be invoked to constrain the results
(Datta 1995; Kogan 1996).
This relation is no longer valid for high-field shot noise
(Green and Das 2000{\it b}).
Once pushed out of the weak-field limit,
drift-diffusive theories face a complex
problem of self-consistency in estimating $D$.
The problem is highly non-linear and ill-controlled.
In the standard Boltzmann approach, there is no {\it a priori}
distinction between `drift' and `diffusion'.
The exact non-equilibrium kinetic equation will always
be linear in the basic electron-hole pair fluctuations.

\item{\bf Langevin sources}.
Gomila and Reggiani follow the popular stratagem of generating
{\it single-particle} fluctuations only, in a drift-diffusive setting.
This is done by adding {\it ad hoc} stochastic source terms
(Langevin's Ansatz) to the equation of motion (Kogan 1996).
The low-order correlators within the phenomenology
must be set by hand to meet the presumed constraint
of Einstein's relation.
The status of this low-field stochastic Ansatz is unclear
in high-field situations; it is certainly no clearer for carrier
populations with strong internal interactions.
Such ambiguities arise simply because the imposition
of extraneous, stochastic current sources has no physical
or logical basis in the microscopics of an {\it internally}
correlated system (van Kampen 1981).
Reliable kinetic descriptions of noise can be set up
with no appeal at all to Langevin sources of the kind adopted for
the Gomila-Reggiani model.
This is as true of non-degenerate noise
%\cite{kormay}
(Korman and Mayergoyz 1996)
as it is for the semi-classical picture of
electron-hole polarisation fluctuations in metallic systems
%\cite{gd01}.
(Green and Das 2000{\it b}).

\item{\bf $T$-dependence}.
Equations (12) and (16) of Gomila and Reggiani (2000) both give
an overall scaling of their $S_{\rm shot}$ with Johnson-Nyquist noise
$S_0$, and hence with temperature.
The shot noise of Eq. (16) in particular
will then be manifestly $T$-dependent, unlike
true shot noise, unless there is a counterbalancing thermal
denominator in the non-linear term giving the
shot-noise contribution. Such a factor will cancel the
thermal dependence introduced through $S_0$.
While this is likely to be so in the classical
high-$T$ limit, where high-field excess noise has
little dependence on $T$
(Green and Das 1998{\it a}, 2000{\it b}),
it is not clearly so in the degenerate regime of their model.
In that limit, the explicit temperature behaviour of the relevant
parameters $L_D$ and $I_R$ is not given.
Unless that behaviour is known, one cannot say whether
the theory of Gomila and Reggiani recovers
{\it true temperature-independent shot noise}
in bulk metallic wires.
A direct check of Eqs. (12) and (16) yields no countervailing
factor to undo the $T$-dependence entering through $S_0$.
Hence $S_{\rm shot}$ must scale with $T$ in the Gomila-Reggiani
model at strong carrier degeneracy.
\end{itemize}

\section{Summary}

We have reviewed some prior results for shot noise
in degenerate conductors. We conclude that the anomalous
recovery of robust shot noise at bulk scales
(where it is normally extinguished by inelastic
scattering) depends
on pushing the system to large enough currents. In such a
strongly non-equilibrium limit, the average transit time of a
carrier is well given by $eN/I$. When this becomes less than the
typical scattering time, carriers are ballistic and the
high-field shot noise reaches its ideal Schottky value of $2eI$.

A simple but {\it strictly} kinetic model, informed by
a time-of-flight understanding of shot noise, gives a consistent
microscopic picture of such noise. The model emphasises plain inelastic
scattering in a strongly driven conductor, rather than more subtle
and higher-order field effects. On the basis of its clear results,
it suggests that Coulomb-fluctuation corrections need not be fundamental
to the physics of shot noise in long conductors. Furthermore, recovery
of the noise at high currents is not sensitive to quantum
statistics. This is because the energy scale for transport
will eventually outstrip even a large Fermi energy.

We have demonstrated the resurgence of true {\it temperature-independent}
shot noise at high currents, even in long thin 3D samples. For a set
level of the current, shot noise should certainly be much stronger
in samples that are relatively wider as well as shorter.
With this prediction, namely the inhibition of high-field shot noise in a
constricted geometry, we advance an altogether different and major
opportunity for new experiments.

We share one common idea with the approach to shot noise proposed
by Gomila and Reggiani. It is the importance of (necessarily discrete)
number fluctuations as generators of shot noise. This departs
from thermal noise, whose character derives from distributed
and continuous random changes in the carriers' free energy.
In a real sense, we are brought right back to basic
thermodynamics. That is because thermal and shot-noise fluctuations
are echoes of the thermodynamic conjugacy of
the equilibrium variations $\delta \mu/k_{\rm B}T$
and $\delta \ln N$.

Aside from the crucial difference over the significance
of Coulomb corrections in the resurgence of bulk shot noise,
there are major differences of method between
Green and Das (1998{\it a})
and
Gomila and Reggiani (2000).
The main one is our systematic adherence to strict
Boltzmannian kinetics and Fermi-liquid theory
(Green and Das 2000{\it b}).
Decidedly, this sets our investigations
apart from each one of the drift-diffusive
(or so-called Boltzmann-Langevin) works
to be found in the noise literature.
That includes Gomila and Reggiani's.

Boltzmann-Langevin phenomenology relies
(a) on fictitious stochastic sources -- said to
generate the individual, one-body current fluctuations -- and
(b) on essentially classical diffusion to evolve such
single-particle objects.
Neither (a) nor, worse, (b) makes any identifiable connection
with the {\it electron-hole pair symmetry} that is essential
to the actual make-up of the microscopic correlations.
It is remarkable that this electron-hole asymmetry
within diffusively driven transport models
(B\"uttiker 1986) persists all the way up to
the macroscopic scale of the device leads (Fenton 1994).
Because this unbalanced behaviour is built right into the
asymptotic boundary conditions, it is intrinsic to all
drift-diffusive descriptions. They cannot be rid of it.

Each of the assumptions above is equally unfounded when it comes
to charged Fermi liquids at the level of microscopic many-body physics
(Green and Das 2000{\it b}).
The price of their apparent phenomenological simplicity is nothing
less than an unphysical electron-hole asymmetry. The consequence
of that is the failure of drift-diffusive fluctuations
to recover the correct electronic compressibility.
Hence they also fail to meet a most basic condition:
metals cannot sustain inhomogeneous electric fields
beyond the Thomas-Fermi screening length.
This rule is violated by {\it every} diffusively based noise model
(Das and Green 2000).

Langevin stochastics and pseudo-classical diffusion
fail to conform to orthodox quantum kinetics for noise in
charge transport, despite frequent claims to the contrary
(Kogan 1996).
Such schemes are deeply foreign to the real
nature of a degenerate, polarisable electron plasma.
There, electron-hole pair dynamics, the conservation laws,
and the sum rules are utterly central to the physics
(Pines and Nozi\`eres 1966).
The drift-diffusive theories' demonstrable violation of
the sum rules, and neglect of the conservation laws
embodied in those rules,
provides the plainest evidence of non-conformity
(Green and Das 2000{\it a,b}).

In passing we have called attention, once again, to the
problematic status of the `smooth crossover' proclaimed by
drift-diffusive phenomenology. We suggest that the time is
ripe for a thorough microscopic reassessment of this
largely intuitive construct, and most of all for a renewed search
for decisive experimental tests of it.

New information on the `smooth crossover' can be expected in two
experimental contexts. The first is in
the low-field noise signal from quantum-confined devices
%\cite{upon,gdii}.
(Green and Das 2000{\it a,c}).
The second is the high-field regime as indicated by us
%\cite{gdcond,gdii}
(Green and Das 1998{\it a}, 2000{\it b,c}),
by Naveh to some extent (Naveh 1998)
and, latterly, by Gomila and Reggiani (2000).
In all situations it is important to have a unifying microscopic 
description, equipped to cover {\it all} of the many facets
of real shot noise. This would be the only way to make the most
of any fresh experimental knowledge.

\section*{references}

Blanter, Ya. M., and B\"uttiker, M. (2000).
{\it Phys. Rept.} {\bf 336}, 1.

B\"uttiker. M. (1986). {\it Phys. Rev. Lett.} {\bf 57}, 1761.

Das, M. P., and Green, F. (2000). `Proceedings of the 23rd International
Workshop on Condensed Matter Theories, Ithaca, Greece',
ed. G. S. Anagnostatos
(Nova Science: New York) in press;
Das, M. P., and Green, F. (1999). Preprint cond-mat/9910183.

Datta, S. (1995). `Electronic Transport in Mesoscopic Systems'
(Cambridge University Press: Cambridge).

Fenton, E. W. (1994). {\it Superlattices and Microstructures}
{\bf 16}, 87.

Gantsevich S., V., Gurevich, V. L., and Katilius, R.
(1979). {\it Nuovo Cimento} {\bf 2}, 1.

Gillespie, D. T. (2000).
{\it J. Phys.: Cond. Matter} {\bf 12}, 4195..

Gomila, G., and Reggiani, L. (2000). {\it Phys. Rev.} B {\bf 62}, 8068.
See also Preprint cond-mat/0005094.

Green, F.,  and Das, M. P. (1998{\it a}). Preprint cond-mat/9809339.
(CSIRO-RPP3911: unpublished.)

Green, F.,  and Das, M. P. (1998{\it b}).
`Recent Progress in Many-Body Theories',
ed. D. Neilson and R. F. Bishop, p. 102 (World Scientific: Singapore).
See also Green, F, and Das, M. P. (1997). Preprint cond-mat/9709142.

Green, F.,  and Das, M. P. (2000{\it a}). `Proceedings of the Second
International Conference on Unsolved Problems of Noise
and Fluctuations (UPoN'99)', ed. D. Abbott and L. B. Kish
AIP {\bf 511}, pp 422-33 (American Institute of Physics: New York).
For a similar discussion see
Green, F., and Das, M. P. (1999). Preprint cond-mat/9905086.
 
Green, F., and Das, M. P. (2000{\it b}).
{\it J. Phys.: Cond. Matter} {\bf 12}, 5233.
See also Green, F., and Das, M. P. (2000). Preprint cond-mat/0001412.

Green, F., and Das, M. P. (2000{\it c}).
{\it J. Phys.: Cond. Matter} {\bf 12}, 5251.
See also Green, F., and Das, M. P. (1999). Preprint cond-mat/9911251.

Kadanoff, L. P., and Baym G. (1962). `Quantum Statistical Mechanics'
(W A Benjamin, Reading, Massachusetts).

Kogan, Sh. M. (1996). `Electronic Noise and Fluctuations in Solids'
(Cambridge University Press: Cambridge).

Korman, C. E.,  and Mayergoyz, I. D. (1996). {\it Phys. Rev.} B
{\bf 54}, 17620.

Naveh, Y. (1998). Preprint cond-mat/9806348.

Pines, D., and Nozi\`eres, P. (1966). `The Theory of Quantum Liquids'
(Benjamin, New York).

van Kampen, N. G. (1981). `Stochastic Processes in Physics
and Chemistry' (North-Holland, Amsterdam), pp 246-52.

%\end{references}

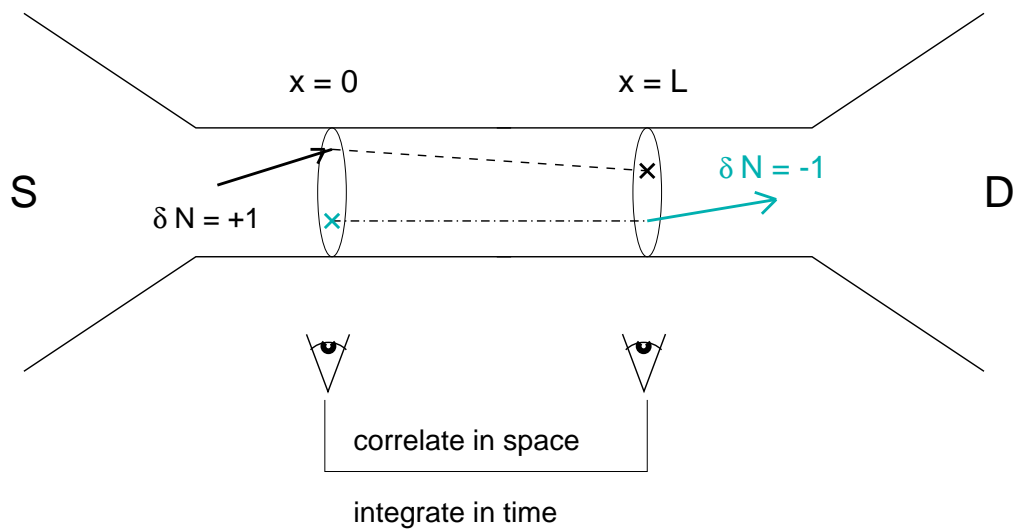
\begin{figure}
\caption{
An operational definition of shot noise. A carrier enters/exits
the sample at a random time, augmenting/depleting the carrier
population by one. The remotely generated disturbance is
also observed at the location of the complementary interface.
This correlation of the stimulus and
its response is equivalent to a time-of-flight measurement,
pairing the carrier states at each observation point. Shot noise
is the sum of these stochastically distributed events. 
}
\label{fig1}
\end{figure}

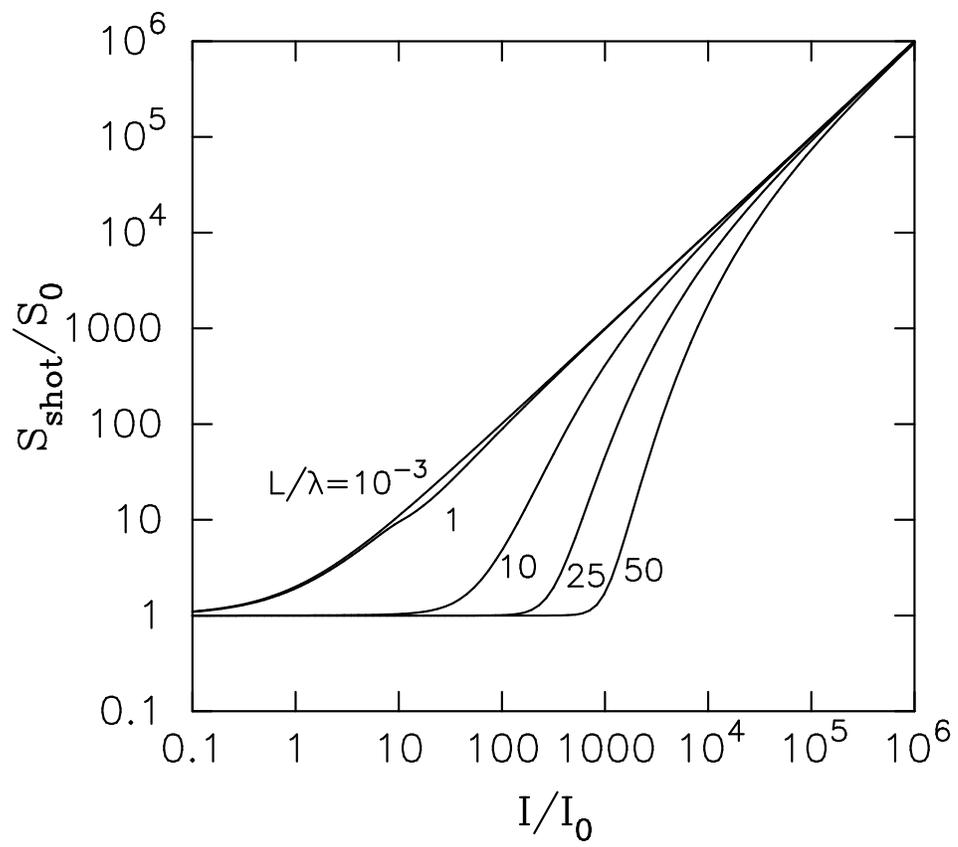
\begin{figure}
\caption{
Sum of thermal and shot noise in a one-dimensional degenerate
wire with $k_{\rm B}T/\varepsilon_{\rm F} = 0.10$. Noise is scaled to the
Johnson-Nyquist value $S_0 = 4Gk_{\rm B}T$,
the current to $I_0 = 2Gk_{\rm B}T/e$.
All the following figures are structured similarly. The curves are indexed,
in descending order, by the ratio of sample length to
Fermi mean free path:
$L/\lambda = 0.001, 1, 10, 25, 50$.
Note the strong attenuation of low-field shot noise for longer
samples. At high enough currents, the shot noise always recovers.
}
\label{fig2}
\end{figure}

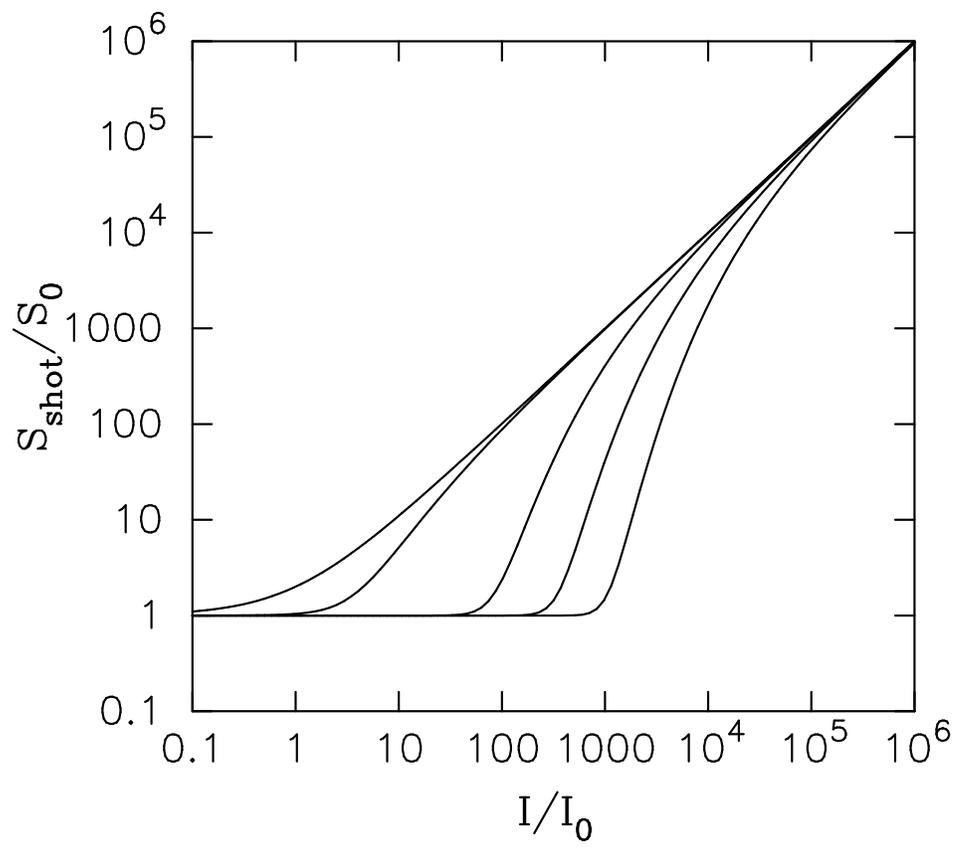
\begin{figure}
\caption{
As for Fig. 2, but for classical carriers at the
same density and temperature, and with the
same set of physical conductor lengths $L$. The second
curve is much reduced in the low-current region where,
in Fig. 2, Pauli blocking inhibits
inelastic scattering and sustains the low-current shot noise.
Note the congruence of the high-current part of the curves with
those of Fig.2. This shows that true high-field shot noise
loses its sensitivity to statistics when the energy
scale for transport substantially exceeds the Fermi energy.
}
\label{fig3}
\end{figure}

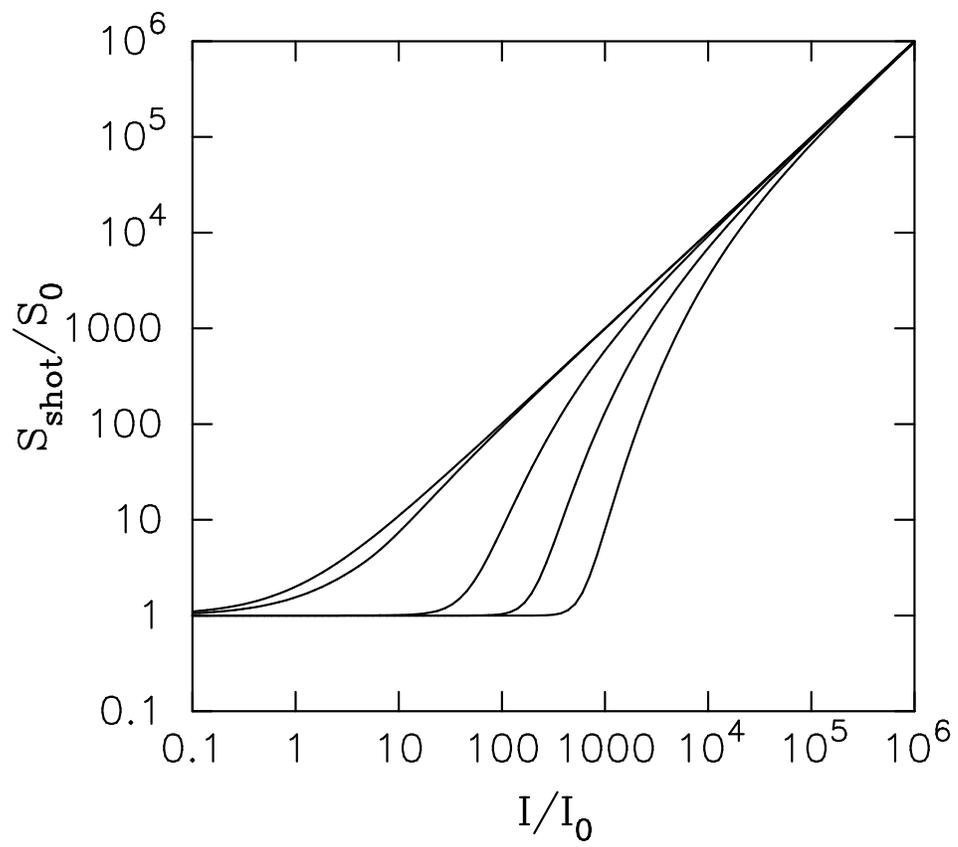
\begin{figure}
\caption{
Three-dimensional (3D) metallic wire of macroscopic thickness. This
is our closest example to that of Fig.1 in Gomila and Reggiani.
The overall concordance of the two results is striking.
}
\label{fig4}
\end{figure}

\begin{figure}
\caption{
(a) A narrow 3D metallic wire, with radius $R = 0.3\lambda$.
Note the large
loss in shot-noise level, and its recovery at higher currents
than for Fig. 4, the large-radius example. 
(b) A very narrow wire, $R = 0.05 \lambda$. The attenuation
is severe, with recovery only at the highest values of current.
}
\label{fig5}
\end{figure}

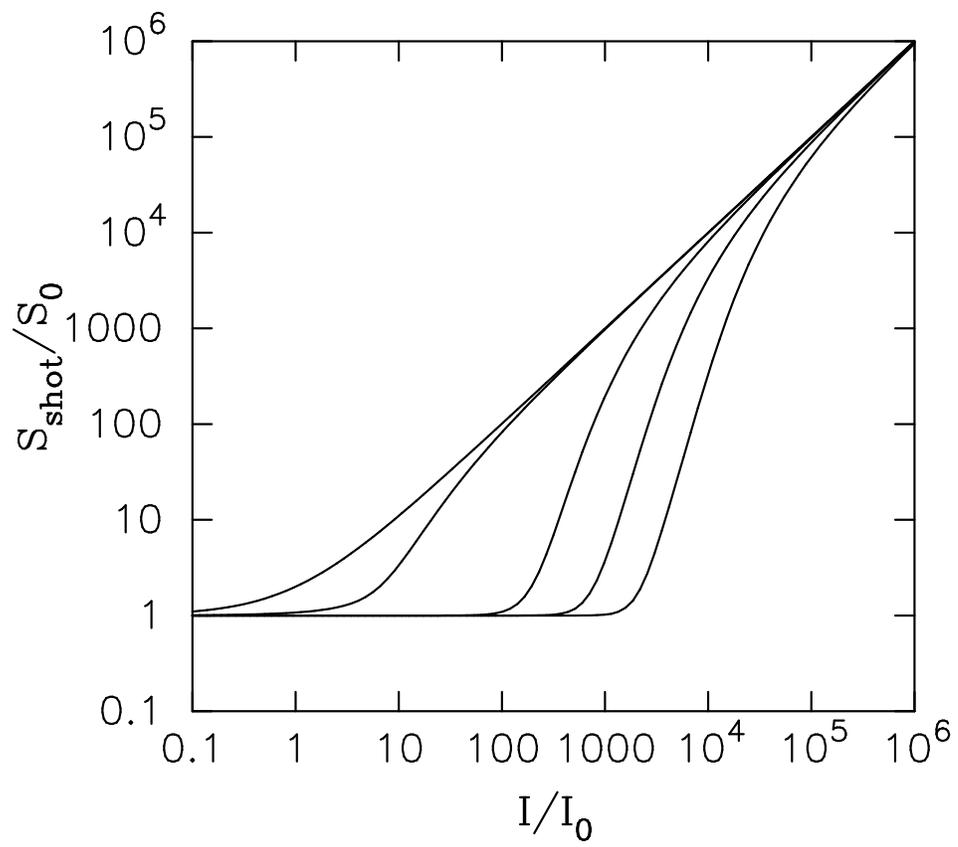
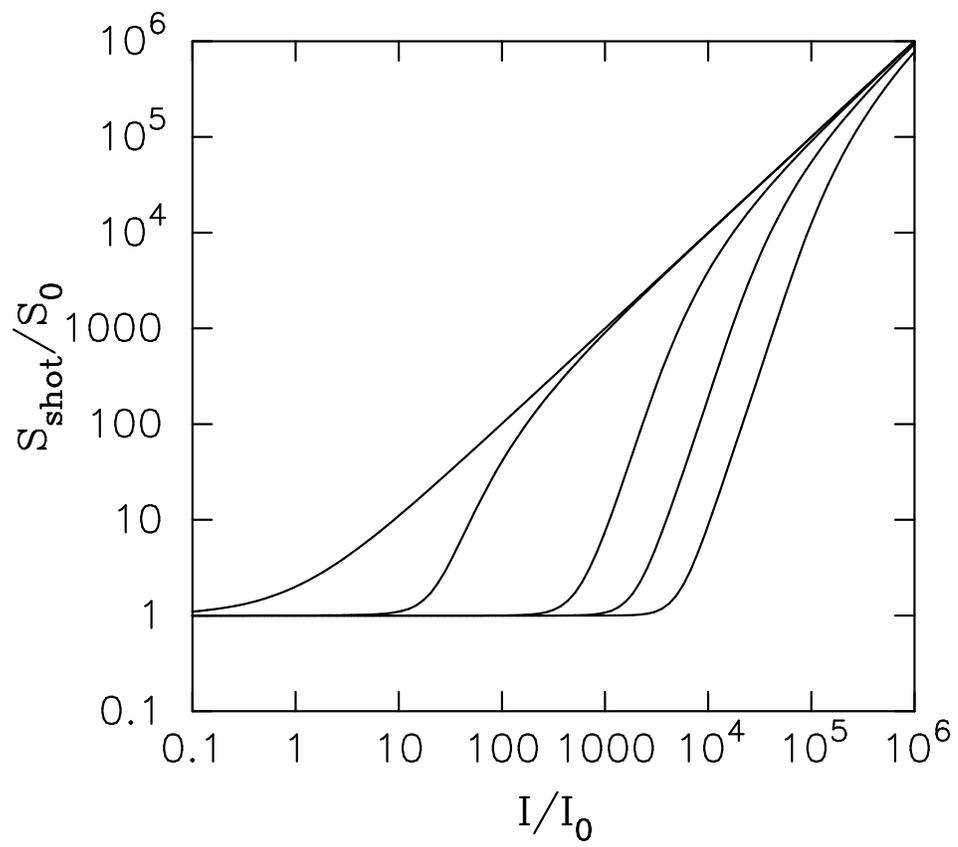
\end{document}